\documentclass[aps,pra,superscriptaddress,amsmath,amssymb,preprintnumbers,floatfix,showpacs,11pt]{revtex4}

\usepackage{amssymb}
\usepackage{epsfig}

\begin{document}
\title{ Revival-collapse phenomenon in the fluctuations of quadrature field
components of the multiphoton Jaynes-Cummings model }
\author{ Faisal A. A. El-Orany  }
 \affiliation{ Department of Mathematics  and Computer Science,
Faculty of Natural Science, Suez Canal University,
 Ismailia, Egypt}

\date{\today}

\begin{abstract}
In this paper we consider a system consisting of a two-level atom,
initially prepared in a coherent superposition of upper and lower
levels, interacting with a radiation field prepared in generalized
quantum states in the framework of multiphoton Jaynes-Cummings
model. For this system we show that there is  a class of states
for which the fluctuation factors can exhibit revival-collapse
phenomenon (RCP) similar to that exhibited in the corresponding
atomic inversion. This is shown not only for normal fluctuations
but also for amplitude-squared fluctuations. Furthermore, apart
from this class of states
 we generally demonstrate that the fluctuation factors associated with three-photon transition
can provide RCP similar to that occurring in the atomic inversion
of the one-photon transition. These are novel results and their
consequence is that RCP occurred in the atomic inversion can be
measured via a  homodyne detector. Furthermore, we discuss the
influence of the atomic relative phases on such phenomenon.

\end{abstract}

  \maketitle

\section{Introduction}
 Interaction between the radiation field and  matter is an important
 topic in modern physics.
One of the most important systems, which describing well the field-matter
interaction is the  Jaynes-Cummings model (JCM). JCM consists
 of a single two-state system interacting with
a single quantized radiation field mode \cite{jay1}.
Furthermore, JCM  has  become experimentally realizable with
the Rydberg atoms in high-$Q$ microwave cavities (, e.g., see \cite{remp}).
Moreover, JCM is exactly solvable under the rotating wave approximation  and many of
interesting phenomena have been observed.
The most important phenomenon is the behavior of the population inversion
where instead of displaying  steady Rabi oscillations in the
case of a classical field coupled to the atom \cite{allen}, there is an initial
collapse of these oscillations followed by regular revivals that slowly
become broader and eventually overlap \cite{eber}.
In fact, the revival-collapse phenomenon (RCP) of the atomic inversion
 is a pure quantum mechanical effect  having its
origin in the granular structure of the photon-number distribution of the
initial field \cite{zaber1}. The systematic and characteristics
of RCP for JCM have been analyzed in details in \cite{eber}.
Moreover, it has been shown that the envelope of each revival is a readout
of the photon distribution, in particular, for the states whose photon-number
 distributions are slowly varying \cite{fleisch}.
It is worth mentioning that observation of RCP has been
performed using the one-atom mazer \cite{remp}, which is more sophisticated than the
dynamics of the JCM.

On the other hand, quadrature fluctuations of the field components
are important quantities in
quantum optics, which
can be measured by a homodyne detection in
which the signal is superimposed on a strong coherent beam of the local
oscillator. The question we would like to address here: Can
the quadrature fluctuations of the multiphoton JCM include information
on RCP of the atomic inversion?
If it is so then
RCP can be detected via a homodyne detector. In other words,
 the quadrature fluctuations as well as atomic inversion of the JCM
can be measured by means of   one device. In this case the scheme
will be simple, involving one beam splitter and a reference field in a coherent
state.
In the present paper we show that such behavior can be occurred.
Specifically, we show that the radiation-field
fluctuation (i.e. squeezing) factors of
the cubic JCM can carry information on the atomic inversion of the
standard JCM (, i.e. JCM which involves one photon for making atomic transition) for the same initial states.
Moreover,  we show that there is a class of states whose fluctuation factors
can include explicitly information on RCP.
Furthermore, we demonstrate that  such phenomenon can occur in the higher-order
fluctuation, e.g. amplitude-squared fluctuations, too. In fact, these are
 novel results and they may be useful for
experimentalists. We have to stress that in this paper we are not
looking for squeezing of the JCM, which has been intensively
studied by several authors (, e.g., see \cite{meys,kim2}).
Nevertheless, we  look at the occurrence of the RCP in the
fluctuation factors. This will be investigated in the following
order. In section 2 we give the basic calculations related to the
system under consideration. In sections 3 and 4 we discuss the
occurrence of RCP in the normal fluctuations and amplitude-squared
fluctuations, respectively. The results are summarized in section
5.

\section{General considerations }
In this section we give both the explicit form for the hamiltonian
of the system under consideration and
the basic calculations related to such system.
The system considered in this paper is
 the multiphoton resonance interaction of a single-mode field
with a two-level atom, which is described by the $m$th-photon JCM.
The effective hamiltonian controlling the system in the rotating wave
approximation (RWA) is \cite{mir}
\begin{equation}
\frac{\hat{H}}{\hbar}=\omega_{0}\hat{a}^{\dagger}\hat{a}+
\omega_{a}\hat{\sigma}_{z}+\lambda (\hat{a}^{m}\hat{\sigma}_{+} +
\hat{a}^{\dagger m}\hat{\sigma}_{-}),
 \label{6}
\end{equation}
where $\hat{\sigma}_{\pm}$ and $\hat{\sigma}_{z}$ are the Pauli spin
operators;
$\omega_{0}$ and $\omega_{a}$ are the frequencies of
cavity mode and the atomic transitions, respectively;  $\lambda$ is the atom-field coupling
constant and $m$ is the  number of photons involved in the atomic transition.
Defining two new operators as
\begin{equation}
\hat{C}_{1}=\omega_{0}\hat{a}^{\dagger}\hat{a}+\omega_{a}\hat{\sigma}_{z},\qquad
\hat{C}_{2}=\lambda (\hat{a}^{m}\hat{\sigma}_{+}+\hat{a}^{\dagger m}\hat{\sigma}_{-}). \label{7}
\end{equation}
In the exact resonance case (, i.e. $\omega_{a}=m\omega_{0}$)
it is easy to prove that $\hat{C}_{1}$ and $\hat{C}_{2}$ are constants
of motion and also they  commute with each other.
This fact makes that the evolution of the mean-photon number and the atomic
inversion of the system include typical information on each other.
In the interaction
picture  the unitary evolution operator takes the form
\begin{eqnarray}
\begin{array}{rl}
\hat{U}_{I}(T,0)=\exp(-i\frac{T}{\lambda}\hat{C}_{2})
\\
\\
=\cos (T\hat{D})-i\frac{\sin (T\hat{D})}{\lambda\hat{D}}\hat{C}_{2},\label{8}
\end{array}
\end{eqnarray}
where
\begin{equation}
T=\lambda t,\qquad \hat{D}^{2}=
\hat{a}^{\dagger m}\hat{a}^{m}\hat{\sigma}_{-}\hat{\sigma}_{+}
+\hat{a}^{m}\hat{a}^{\dagger m}\hat{\sigma}_{+}\hat{\sigma}_{-}. \label{9}
\end{equation}
It is worth reminding that $\hat{\sigma}_{\pm}^2=0$.

On the other hand, to keep the analysis quite general, we consider
the field prepared initially in a general pure quantum state describing by
\begin{equation}
|\psi(0)\rangle=\sum\limits_{n=0}^{\infty}C_{n}|kn\rangle, \label{1}
\end{equation}
where $C_{n}$ represent the probability amplitudes for the
state under consideration such that
$\sum\limits_{n=0}^{\infty}|C_{n}|^{2}=1$, and $k$ is a parameter its value will be
 specified in the text.
 Throughout the paper we consider the probability amplitudes $C_{n}$
to be real.
 It is worth mentioning that
when $C_{n}$ represent the probability amplitudes of the
well-known Gluaber coherent state and $k\neq 1$ then (\ref{1})
gives the $k$-photon coherent  states \cite{jex1,jex2}. These
states are obtained from $k$th harmonic generation using
Brandt-Greenberg operators \cite{green}. It has been shown that
such a class  of states can exhibit  amplitude $k$th-power
squeezing
 \cite{jex2}  when they interact with the
nonlinear nonabsorbing medium modeled as an anharmonic oscillator.
We proceed by considering  that the atom is initially  in the coherent
 superposition of the excited and ground states as
\begin{equation}
|\theta,\phi\rangle=\cos\theta |+\rangle+\exp(-i\phi)\sin\theta
|-\rangle,\label{2}
\end{equation}
where $|+\rangle$ and
$|-\rangle$ denote excited and ground atomic states, respectively;
$\theta$
and  $\phi$ are the relative phases between these two atomic states.
Actually, preparing the atom in the coherent superposition
states is important because of its applications to noise quenching by
correlated spontaneous emission \cite{quen1}, quantum beats \cite{quen2},
and noise-free amplification \cite{quen3}.

Now the  initial state of the field-atom system  can be expressed as
\begin{equation}
|\Psi (0)\rangle=|\psi (0)\rangle\bigotimes
 |\theta,\phi\rangle. \label{3}
\end{equation}
Therefore, the  dynamical wave function  of the total system in the interaction
picture is given by
\begin{eqnarray}
\begin{array}{lr}
|\Psi(T)\rangle=\hat{U}_{I}(T,0)|\Psi(0)\rangle\\
\\
=\sum\limits_{n=0}^{\infty}\left[G_{1}(n,T)|+,n\rangle+G_{2}(n,T)|-,n+m\rangle
\right], \label{10}
\end{array}
\end{eqnarray}
where
\begin{eqnarray}
\begin{array}{lr}
G_{1}(n,T)=C_{n}\cos\theta \cos (T\sqrt{h(n,m)})
-i\exp(-i\phi)C_{n+m}\sin\theta \sin (T\sqrt{h(n,m)}),\\
\\
G_{2}(n,T)=\exp(-i\phi)C_{n+m}\sin\theta \cos (T\sqrt{h(n,m)})
-iC_{n}\cos\theta \sin (T\sqrt{h(n,m)}),
\label{11}
\end{array}
\end{eqnarray}
while $h(n,m)=\frac{(n+m)!}{n!}$ and in the course of the calculation
 we have considered $k=1$ (cf. (\ref{1})).
For the future purpose, we derive different moments for the $\hat{a}^{\dagger}$
and $\hat{a}$ associated with  the state (\ref{10}) as
\begin{eqnarray}
\begin{array}{lr}
\langle\hat{a}^{\dagger s_{2}}(T)
\hat{a}^{s_{1}}(T)
\rangle=\sum\limits_{n=0}^{\infty}\left[
G^{*}_{1}(n+s_{2},T)G_{1}(n+s_{1},T)
\frac{\sqrt{(n+s_{1})!(n+s_{2})! }}{n!}\right.\\
\\
+\left.
G^{*}_{2}(n+s_{2},T)G_{2}(n+s_{1},T)
\frac{\sqrt{(n+m+s_{1})!(n+m+s_{2})! }}{(n+m)!}
\right], \label{12}
\end{array}
\end{eqnarray}
where $s_{1}$ and $s_{2}$ are positive integers.
Also the atomic inversion for the dynamical state (\ref{10}) is
\begin{eqnarray}
\begin{array}{lr}
\langle
\sigma_{z}(T)\rangle=\sum\limits_{n=0}^{\infty}\Bigl\{
\left[P(n) \cos^{2}\theta
 -P(n+m)\sin^{2}\theta\right] \cos(2T\sqrt{h(n,m)})
 \\\\
-\sqrt{P(n)P(n+m)} \sin\phi \sin (2\theta)
\sin(2T\sqrt{h(n,m)})\Bigr\},
\label{13} \end{array}
\end{eqnarray}
where $P(n)=C_{n}^{2}$.

We close this section by mentioning that, according to
the lines  given in \cite{Toor}, the use of the
hamiltonian (\ref{6}) is called an effective hamiltonian approach (EHA).
Nevertheless,
the full microscopic hamiltonian approach (FMHA) associated with the system
 can be obtained by considering the hamiltonian, which describes
 the interaction between
 $(m+1)$th-level atom in a cascade configuration
 with the single-mode radiation field in the RWA \cite{Fu}.
Under certain condition the intermediate levels can be canceled out
adiabatically and the
system reduced to that of the two-level atom.
In this case the probability amplitudes of the dynamical wave function
of the system include nontrivial  overall phase depending on the intensity
of the field. This makes  the results associated with FMHA are
completely different from those with EHA,
in particular,  quantities that
depend on the off-diagonal elements of the reduced density
matrix such as the FQFC.
 Alternatively, the
hamiltonian (\ref{6}) can be modified to provide similar information--under certain
conditions--as that of FMHA \cite{Toor}.  This can be achieved by inclusion
 the dynamic Stark shift in (\ref{6}), i.e. including such a term
$-\hat{a}^{\dagger}\hat{a}(\beta_{1}\hat{\sigma}_{+}\hat{\sigma}_{-}+
\beta_{2}\hat{\sigma}_{-}\hat{\sigma}_{+})$ in (\ref{6}) where
$\beta_{1},\beta_{2}$ are dynamic Stark
shift parameters. This technique is called modified effective hamiltonian approach
(MEHA) and for the sake of comparison   we give some details
about it.
For instance, the dynamical state  for the system associated with MEHA in the interaction
picture (considering the initial condition (\ref{3})) is
\begin{equation}
|\tilde{\Psi}(T)\rangle
=\sum\limits_{n=0}^{\infty}\left[\tilde{G}_{1}(n,T)|+,n\rangle+\tilde{G}_{2}(n,T)|-,n+m\rangle
\right], \label{ma2}
\end{equation}
where
\begin{eqnarray}
\begin{array}{lr}
\tilde{G}_{1}(n,T)=\exp(it V_{n})\left\{ C_{n}\cos\theta \cos (t\Omega_{n})\right.
\\\\
+\left.\frac{i}{\Omega_{n}}\left[(n\beta_{1}-V_{n})C_{n}\cos\theta
-\lambda \sqrt{\frac{(n+m)!}{n!}}\exp(-i\phi)C_{n+m}\sin\theta\right]
\sin (t\Omega_{n})\right\},
\\
\\
\tilde{G}_{2}(n,T)=\exp(itV_{n})\left\{
\exp(-i\phi)C_{n+m}\sin\theta\cos(t\Omega_{n})\right.
\\
\\
-\left.\frac{i}{\Omega_{n}}\left[
(V_{n} -(n+m)\beta_{2})
\exp(-i\phi)C_{n+m}\sin\theta
+\lambda \sqrt{\frac{(n+m)!}{n!}}C_{n}\cos\theta\right]
\sin (t\Omega_{n})\right\}
\label{ma3}
\end{array}
\end{eqnarray}

and
\begin{equation}
V_{n}=\frac{1}{2}[n\beta_{1}+(n+m)\beta_{2}], \quad
\Omega_{n}=\frac{1}{2}\left[(n\beta_{1}-(n+m)\beta_{2})^{2}+
4\lambda \frac{(n+m)!}{n!}\right]^{\frac{1}{2}}.
\label{ma4}
\end{equation}
When $m=2$ expressions (\ref{ma3})--(\ref{ma4})
reduce to (40)--(43) in \cite{Toor}. By the way there is a misprint in (41)
of \cite{Toor}
where the term $(n\beta_{1}-V_{n})$ has to be $(V_{n} -(n+2)\beta_{2})$.
Comparison between (\ref{11}) and (\ref{ma3}) shows that involving
 the dynamic Stark shift in the effective hamiltonian makes
the probability amplitudes including nontrivial  overall phase, which depends on the
intensity of the field, as we mentioned above in relation to  FMHA.

Throughout the paper we focus the attention on EHA. To be more specific,
we use expressions (\ref{12}) and (\ref{13}) to make a comparative study
between the behavior of the fluctuation factors and atomic inversion.
Also  we give only some comments on MEHA aiming to show the
differences between EHA and MEHA.
So the discussion is  generally given for EHA, except specifying that it is
related to MEHA.

\section{Revival-collapse phenomenon in  normal fluctuations}
In this section,  we show  that information stored in
$\langle\hat{\sigma}_{z}(T)\rangle$ can be obtained from fluctuation
factors of the second-order (normal) fluctuation. To do so we define
 two quadrature operators as
$\hat{X}=\frac{1}{2}[\hat{a}+\hat{a}^{\dagger}], \quad
\hat{Y}=\frac{1}{2i}[\hat{a}-\hat{a}^{\dagger}]$.  These quadratures
satisfy the commutation rule $[ \hat{X},\hat{Y}] =\frac{i}{2}$ and
thus  the uncertainty relation is
$\langle (\triangle\hat{X}(T))^{2}\rangle
  \langle (\triangle\hat{Y}(T))^{2}\rangle \geq \frac{1}{16}$.
Therefore, the fluctuation factors associated with the quadratures $\hat{X}$
and  $\hat{Y}$, respectively, read
\begin{eqnarray}
\begin{array}{lr}
F_{1}(T)=2\langle (\triangle \hat{X}(T))^{2}\rangle-\frac{1}{2}  \\
\\
= \langle\hat{a}^{\dagger}(T)\hat{a}(T)\rangle+{\rm Re}
\langle\hat{a}^{2}(T)\rangle-2({\rm
Re}\langle\hat{a}(T)\rangle)^{2},
\\
\\
S_{1}(T)=2\langle (\triangle \hat{Y}(T))^{2}\rangle-\frac{1}{2}
\\
\\
= \langle\hat{a}^{\dagger}(T)\hat{a}(T)\rangle-{\rm Re}
\langle\hat{a}^{2}(T)\rangle-2({\rm
Im}\langle\hat{a}(T)\rangle)^{2}.
\label{14}
\end{array}
\end{eqnarray}
The system is able to yield  normal squeezing when $F_{1}(T)<0$ or
$S_{1}(T)<0$, however, this is not the aim of this paper.
Based on (\ref{14}) we illustrate that there are two approaches, namely,
natural phenomenon and numerical simulation, which can provide RCP in
$F_{1}(T)$ and/or in $S_{1}(T)$.
In the first approach we show that there is
particular class  of states that can naturally exhibit RCP in the fluctuation factors.
Nevertheless, in the second approach we demonstrate that $S_{1}(T)$, for
particular values of   $m$, can exhibit similar behavior
as that of
$\langle\hat{\sigma}_{z}(T)\rangle$ of the standard JCM.
In fact these two approaches are related to two different situations  in
which different terms  dominate the variance of the field amplitude.
To be more specific, for the natural phenomenon
 the origin of RCP in the normal fluctuation is the
 $\langle \hat{a}^{\dagger}(T) \hat{a}(T)\rangle$, however, in the numerical
 simulation approach is the  ${\rm Re}\langle \hat{a}^{2}(T)\rangle$, as
 we will show below.
Furthermore, we investigate the influence of the atomic relative phases on the
occurrence of RCP in the fluctuation factors.
Also we make some comments on the differences between EHA and MEHA related to the
under consideration phenomenon.
These points will be investigated in the following two parts.
\subsection{Natural phenomenon}
This approach is based on the fact that $\hat{C}_{1}$ is a constant
of motion and then the evolution of the
$\langle\hat{a}^{\dagger }(T)\hat{a}(T)\rangle$ and
$\langle\hat{\sigma}_{z}(T)\rangle$ for the same value of $m$
 yield similar behavior.
So that if there are  states for which
\begin{equation}
\langle\hat{a}(T)\rangle=0,\qquad
\langle\hat{a}^{2}(T)\rangle=0,
\label{15}
\end{equation}
simultaneously then the two fluctuation factors  in (\ref{14})
reduce to $\langle\hat{a}^{\dagger }(T)\hat{a}(T)\rangle$.
In other words, $F_{1}(T)$ and/or $S_{1}(T)$ provide
an information on  the atomic inversion.
Now we are looking for such type of states.
For convenience we restrict the analysis to
 $m=1$ and $\theta=0$. The associated quantities with this case
 can be obtained from
(\ref{12}) as
\begin{eqnarray}
\begin{array}{lr}
\langle\hat{a}(T)\rangle
 =\sum\limits_{n=0}^{\infty}
C_{n}C_{n+1}\sqrt{n+1}
\Bigl\{\cos [T\sqrt{h(n,1)}] \cos [T\sqrt{h(n+1,1)}]
 \\
 \\
+\sqrt{\frac{n+2}{n+1}}
\sin [T\sqrt{h(n,1)}] \sin [T\sqrt{h(n+1,1)}]
\Bigr\},\\
\\
\langle\hat{a}^{2}(T)\rangle
 =\sum\limits_{n=0}^{\infty}
C_{n}C_{n+2}\sqrt{(n+1)(n+2)}
\Bigl\{\cos [T\sqrt{h(n,1)}] \cos [T\sqrt{h(n+2,1)}]
 \\
 \\
+\sqrt{\frac{n+3}{n+1}}
\sin [T\sqrt{h(n,1)}] \sin [T\sqrt{h(n+2,1)}]
\Bigr\}.
\end{array}
 \label{16}
 \end{eqnarray}
It is obvious that conditions (\ref{15}) are satisfied simultaneously only when
\begin{equation}
C_{n}C_{n+1}=0,\qquad C_{n}C_{n+2}=0 \label{17}
\end{equation}
and these equalities  can be achieved  for three-photon states, four-photon
states and so on. The $k$-photon coherent states  (cf. (\ref{1})) can play this
role, e.g. when $k=3,4,..$, etc. It is worth mentioning that the properties
 of the three-photon states have been investigated in \cite{gener}.
Further, examples of
the four-photon states are the orthogonal-even, (-odd) coherent states
\cite{lyn} and phased generalized binomial states \cite{fais1}.
Here we shed the light on the behavior of $F_{1}(T)$ of the
JCM against the orthogonal-even coherent states. Their forms
  can be obtained from
(\ref{1}) by setting $k=1$ and replacing the probability amplitudes $C_{n}$ by
\begin{equation}
C_{2n}=B\frac{\alpha^{2n}}
{\sqrt{(2n)!}}[1+(-1)^{n}], \label{18}
\end{equation}
where $B$ is the normalization constant having the form
\begin{equation}
B=[2\cosh |\alpha|^{2}+2\cos|\alpha|^{2}]^{-\frac{1}{2}}.\label{19}
\end{equation}
Such type of states have been investigated in \cite{lyn} showing
that they cannot exhibit second-order squeezing, whereas
near-optimal simultaneous-quadrature fourth-order squeezing can be
obtained. Also they can be generated using conditional-measurement
technique \cite{cm1,jex}. Fig. 1 has been plotted for $F_{1}(T)$
of the EHA with the field initially in orthogonal-even coherent
states for given values of the parameters. From this figure it is
clear that the RCP is established. In fact, the revivals are  four
times compared to those of the corresponding  initial coherent
light since orthogonal-even coherent states are a superposition of
four-component coherent states. This leads to
$T_{R}^{(f)}=T_{R}^{(c)}/4$, where $T_{R}^{(c)}$ and $T_{R}^{(f)}$
are the revival times associated with the initial coherent states
and orthogonal-even coherent states, respectively. This fact can
be deduced as follows. For the initial coherent light the revivals
occur by estimating the time that neighbor terms in the sums are
in phase (for $\bar{n}= \sqrt{\langle \hat{n}(0)\rangle}$, where
$\hat{n}(0)=\hat{a}^{\dagger}(0)\hat{a}(0)$):
\begin{equation}
2 T^{(c)}_{R}\left[ \sqrt{\langle \hat{n}(0)\rangle+1}
-\sqrt{\langle \hat{n}(0)\rangle}\right]\simeq 2\pi, \label{20}
\end{equation}
Nevertheless, orthogonal-even coherent states are four-photon state
and thus the difference in phase
of two (non-zero) neighbor terms will be

\begin{equation}
2 T^{(f)}_{R}\left[ \sqrt{\langle \hat{n}(0)\rangle+4}
-\sqrt{\langle \hat{n}(0)\rangle}\right]\simeq 2\pi. \label{21}
\end{equation}
Expressions (\ref{20}) and (\ref{21}) lead to

\begin{equation}
T^{(c)}_{R}\simeq 2\pi \sqrt{\langle \hat{n}(0)\rangle}, \qquad
T^{(f)}_{R}\simeq \frac{\pi}{2} \sqrt{\langle \hat{n}(0)\rangle}.
\label{22}
\end{equation}
This means  that $ T^{(f)}_{R}=T^{(c)}_{R}/4$.
\begin{figure}
 \includegraphics[width=.60\linewidth]{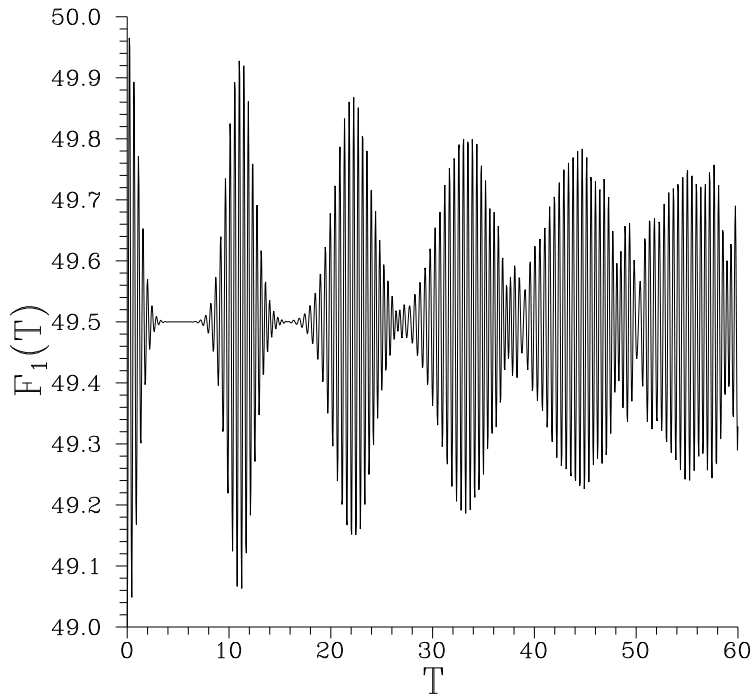}
\caption{ The fluctuation factor $F_{1}(T)$ of the standard JCM
against the scaled time $T$ when the optical cavity field
initially prepared in orthogonal-even coherent states and the atom
is in the excited atomic state for $|\alpha|=7$.}
\end{figure}

The influence of the atomic relative phases on the behavior of
$F_{1}(T)$ for the present approach can be investigated as follows.
As is well known--for the standard JCM and for certain choice of the
atomic phases (, i.e. for $\theta$ and $\phi$)--that
"coherent trapping" occurs  \cite{zaheer}.
Actually, similar conclusion can be given here, i.e. the interaction
has a little effect on $F_{1}(T)$.
For example,
for orthogonal-even coherent states  this can occur when
$\theta=\pi/4, \phi=0$ and  $m=4$.
 The origin in taking $m=4$ is
quite obvious from (\ref{13}), where  atomic trapping occurs when
$\langle \hat{\sigma}_{z}(T)\rangle\simeq 0$
(or in the language of the present approach when $F_{1}(T)\simeq \langle\hat{n}(0)\rangle$), i.e.
\begin{equation}
P(n)-P(m+n)\simeq 0. \label{23a}
\end{equation}
Expression (\ref{23a}) leads to $P(n)\simeq P(n+m)$, i.e. the two successive
non-zero values of the photon-number distribution should be comparable.
This  occurs when $m$ equals  to  the parity of
  the initial  state of the optical cavity field.
More illustratively, atomic trapping for $m$th JCM with optical cavity
field prepared  initially in, e.g., single-, two-, three-  and
four-photon states  occurs only when $m=1,2,3$ and $4$, respectively.

We close  this part by the following remark.
For the natural phenomenon approach EHA and MEHA provide almost similar behavior in relation to the RCP in, e.g., $F_{1}(T)$. In this case the nonvanishing term (, i.e. the mean-photon number) depends only on
the  diagonal elements of the density matrix and then the intensity-dependent
phases in MEHA are canceled out.
We should point out that  the RCP can occur in the fluctuation factors for strong-intensity regime
$ \langle \hat{n}(0)\rangle>>1$, which is the same condition for
EHA and MEHA to provide similar behavior \cite{Toor}.

\subsection{Numerical simulation}
In this part we discuss the possibility to obtain RCP from the second-order
fluctuation factors of the $m$th ($m>2$) JCM similar to that of
$\langle \hat{\sigma}_{z}(T)\rangle$ of the standard JCM, which will be
 denoted by $\langle \hat{\sigma}_{z}(T)\rangle_{m=1}$.
We assume that the initial states are not those for which  the natural
phenomenon can occur.
Careful examination of (\ref{14}) shows that  RCP can occur in $F_{1}(T)$
(or $S_{1}(T)$)  provided that the values of
${\rm Re} \langle \hat{a}(T)\rangle$
(or ${\rm Im} \langle \hat{a}(T)\rangle$) are approximately zero in the
course of the interaction since these quantities are squared and then they
spoil RCP (if it exists).
On the other hand, for $m>2$,
$\langle\hat{a}^{\dagger}(T)\hat{a}(T)\rangle$ exhibits chaotic behavior
(see Fig. 3(a) given below).
Therefore, under these circumstances, if  $S_{1}(T)$, say, can exhibit
RCP then  the origin is in ${\rm Re} \langle \hat{a}^{2}(T)\rangle$.
 For this reason we compare the form of
${\rm Re} \langle \hat{a}^{2}(T)\rangle$
with that of $\langle \hat{\sigma}_{z}(T)\rangle_{m=1}$
for optical cavity
field  initially in coherent states (with real probability amplitudes)
and the atom  in the atomic excited state. The aim of such comparison
is two-fold: (i) To find the exact values of the  number of photons
involved in the atomic transition, i.e. $m$, for
which such phenomenon can occur. (ii) To explore the form of the modified
fluctuation factor, which can include typical information on the behavior
of the $\langle \hat{\sigma}_{z}(T)\rangle_{m=1}$.
Now from (\ref{12}) we arrive at

\begin{eqnarray}
\begin{array}{lr}
\langle
\hat{a}^{2}(T)\rangle=\langle \hat{n}(0)\rangle
\sum\limits_{n=0}^{\infty}P(n)\Bigl[\{\sqrt{
\frac{(n+m+2)(n+m+1)}{(n+2)(n+1)}}\}
\sin (T\sqrt{h(n,m)})\sin (T\sqrt{h(n+2,m)})\\
\\
 +\cos (T\sqrt{h(n,m)})\cos (T\sqrt{h(n+2,m)})\Bigr], \label{23}
\end{array}
\end{eqnarray}
where $P(n)$ is the photon-number distribution for the coherent light and
$\langle \hat{n}(0)\rangle=|\alpha|^{2}$.
We treat the problem in a strong-intensity regime when $m$ is finite.
In this case   the terms contribute effectively to
the summation in (\ref{23}) are those for which  $\alpha^{2}\simeq n$.
 Therefore,  the square root included in the curly brackets
in (\ref{23}) tends to unity and thus reads
\begin{equation}
\langle
\hat{a}^{2}(T)\rangle=\langle \hat{n}(0)\rangle\sum\limits_{n=0}^{\infty}P(n)
\cos [T (\sqrt{h(n+2,m)}-\sqrt{h(n,m)})]. \label{24}
\end{equation}
\begin{figure}
  \includegraphics[width=.60\linewidth]{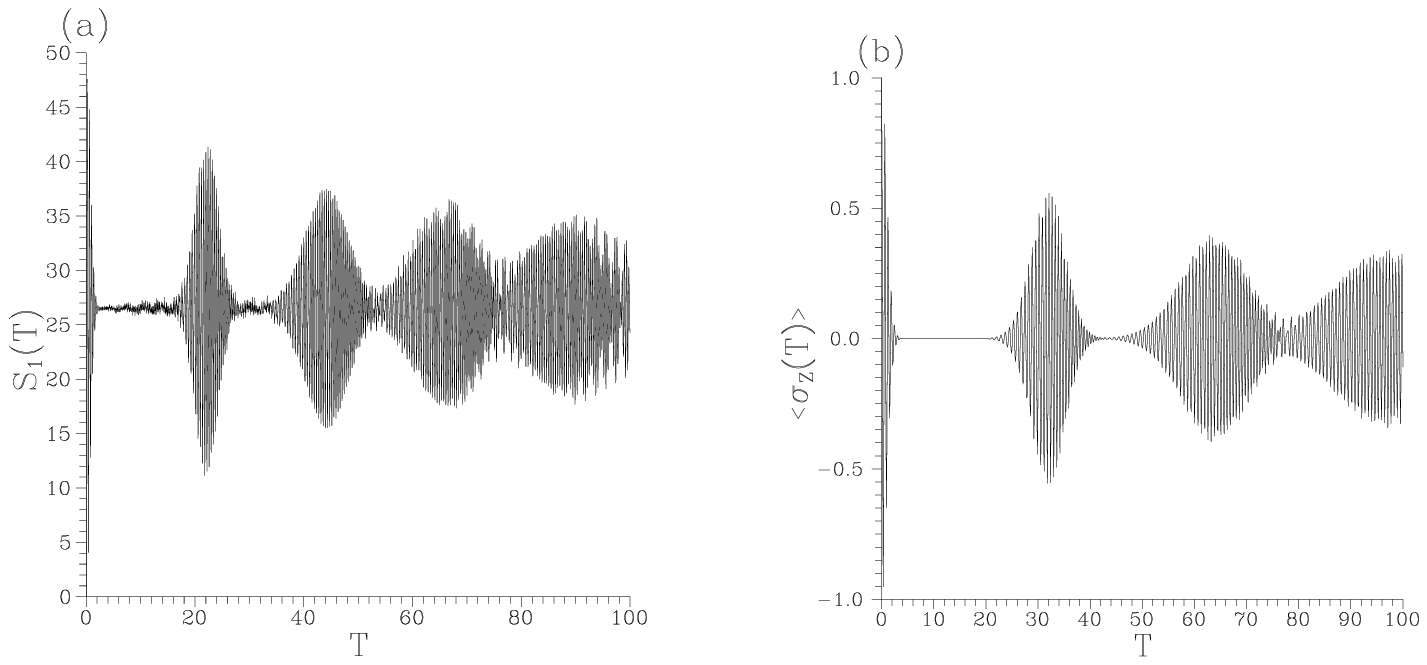}
\caption{ The fluctuation factor $S_{1}(T)$ of the JCM for $m=3$
(a) and the atomic inversion for $m=1$ (b) against the scaled time
$T$ when the field prepared initially in the coherent state with
$|\alpha|=5$ and the atom in the atomic excited state $\theta=0$.}
\end{figure}

On the other hand, the  corresponding atomic inversion of the standard JCM is
\begin{equation}
\langle \hat{\sigma}_{z}(T)\rangle=
\sum\limits_{n=0}^{\infty}P(n)\cos(2T\sqrt{n+1}).\label{25}
\end{equation}
Apart from the constant quantity $\langle \hat{n}(0)\rangle$
in (\ref{24}), expressions (\ref{24})  and (\ref{25})
yield similar behavior provided that the arguments of the $\cos(.)$
 are comparable.
Therefore, we adopt the following  proportionality factor
\begin{equation}
f(n)=
\frac{\sqrt{h(n+2,m)}-\sqrt{h(n,m)}}{2\sqrt{n+1}}.
\label{26}
\end{equation}
After straightforward  calculation (\ref{26}) takes the form
\begin{equation}
f(n)=
\frac{n^{\frac{m-3}{2}}\left[2m+\frac{m}{n}(m+3)\right]\sqrt{\prod\limits_{j=0}^{m-1}(1+\frac{m-j}{n})}
}{2\sqrt{1+\frac{1}{n}}
[\sqrt{(1+\frac{m+1}{n})(1+\frac{m+2}{n})}+
\sqrt{(1+\frac{1}{n})(1+\frac{2}{n})}]}.
\label{27}
\end{equation}
In the strong-intensity regime
 expression (\ref{27}) reduces to
\begin{equation}
f(n)\simeq \frac{m}{2}n^{\frac{m-3}{2}}.\label{28}
\end{equation}
It is evident from (\ref{28}) that the allowed value of $m$ for which
RCP can occur in $S_{1}(T)$ is
 only $m=3$ and thus $f(n)\simeq 3/2$.
The validity of the above facts  has been checked numerically in Figs.
2 and 3.
\begin{figure}
  \includegraphics[width=.60\linewidth]{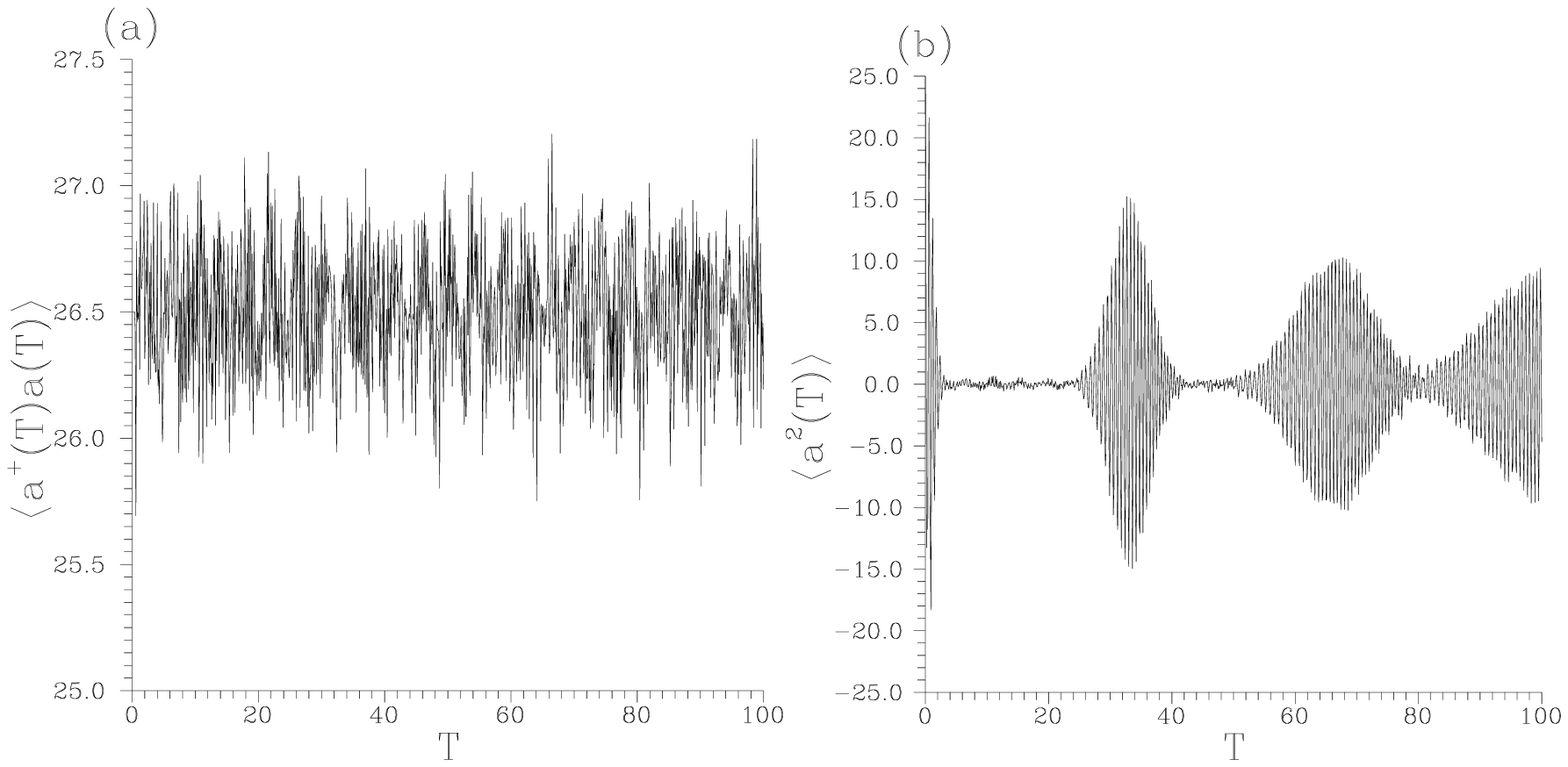}
\caption{ The mean-photon number (a) and the moment $\langle
\hat{a}^{2}(T)\rangle$ (b) against the scaled time $T$ for the
same values of the parameters as those in Figs. 2 but with $m=3$.}
\end{figure}

Figs. 2(a) and (b) have plotted for  $S_{1}(T)$
 and $\langle \hat{\sigma}_{z}(T)\rangle$, respectively, for given values
 of the interaction parameters.
According to above discussion RCP can occur
only in $S_{1}(T)$ (since in this case ${\rm Re}\langle\hat{a}(T)\rangle
\neq 0$ and ${\rm Im}\langle\hat{a}(T)\rangle =0$).
Comparison between Fig. 2(a) and Fig. 2(b) shows
that they roughly exhibit similar behavior in
a sense  that they  revive, collapse, remain quiescent, revive, collapse and so on.
For large interaction time overlapping between successive revivals occurs.
Nevertheless, they include different scales, which we treat   shortly.
Figs. 3(a) and (b) shed the light on the evolution of the
$\langle\hat{a}^{\dagger}(T)\hat{a}(T)\rangle$ and
${\rm Re}\langle \hat{a}^{2}(T)\rangle$, respectively, i.e.
the non-vanishing components
in $S_{1}(T)$. In these figures the values of the parameters
are the same as those in Fig. 2(a).   It is obvious that
${\rm Re}\langle \hat{a}^{2}(T)\rangle$ is responsible
for the occurrence of RCP in $S_{1}(T)$,  as we
have discussed above.
Now within the context  of the above analysis the  rescaled fluctuation factor for
cubic JCM,
which can provide typical information on the $\langle
\hat{\sigma}_{z}(T)\rangle_{m=1}$, is

\begin{equation}
Q_{1}(T)=\frac{S_{1}(\frac{2}{3}T)-\langle\hat{n}(0)\rangle}
{\langle\hat{n}(0)\rangle}. \label{29}
\end{equation}
Fig. 4 is given for $Q_{1}(T)$ that is represented by (\ref{29}) for the
same values of the parameters as those given in Fig. 2(a).
Comparison between Fig. 2(b) and
Fig. 4 is instructive.
Actually, this is a novel result and its consequence is that RCP
 of the $\langle \hat{\sigma}_{z}(T)\rangle_{m=1}$
 can be obtained from the modified fluctuation factor of the
 cubic JCM for the same initial optical cavity
 field.
\begin{figure}
\includegraphics[width=.60\linewidth]{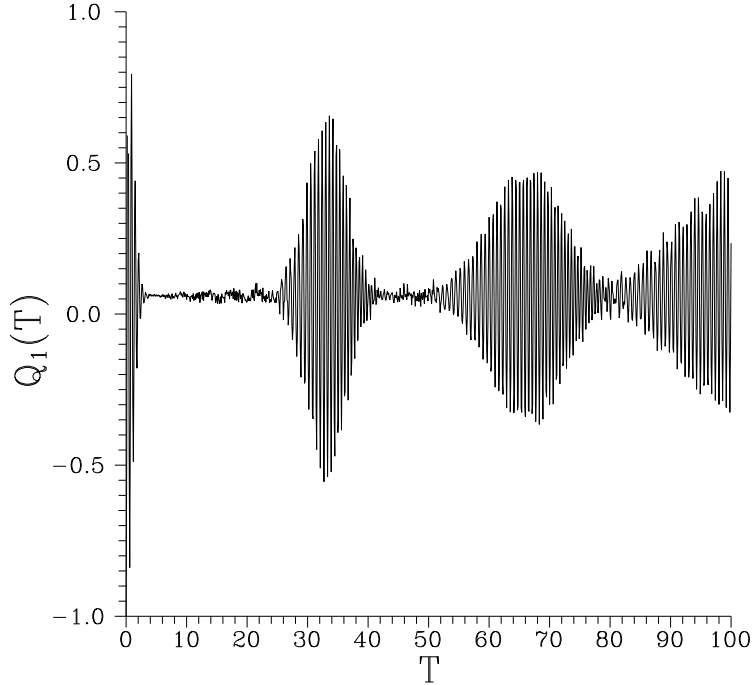}\caption{
The rescaled fluctuation factor $Q_{1}(T)$ given by (28) for the
same values of the parameters as in Fig. 2a.}
\end{figure}

Now we  demonstrate the influence of atomic relative phases
on the behavior of $Q_{1}(T)$. Actually, in contrast to the natural
phenomenon as well as the atomic inversion the rescaled fluctuation factor is insensitive to
the values of the atomic relative phases.
 This fact can easily be recognized, where in the strong-intensity
 regime and for $\theta=\pi/4,\phi=0$, one can show that ${\rm Re}\langle
 \hat{a}^{2}(T)\rangle$ includes such a term
  $[P(n)+P(n+m)]/2$, which cannot be zero for $P(n)\neq 0$.
Therefore, (\ref{29}) yields typical information on the atomic
inversion provided that the atom is either in the excited state or in
the ground state.

From above discussion  generally
 RCP occurred for EHA cannot be established for MEHA since for the
latter  ${Re}\langle \hat{a}(T)\rangle\neq 0$
and ${Im}\langle \hat{a}(T)\rangle\neq 0$ where the probability amplitudes of
  the wave function  include   intensity-phase  dependent (cf. (\ref{ma3})).
Nevertheless, for particular type of states, e.g. parity
states such as the even and odd coherent states,
with strong initial mean-photon number
EHA and MEHA can provide almost similar behavior.
As in this case
$\langle \hat{a}(T)\rangle=0, \frac{\lambda h(n,m)}{\Omega_{n}}\simeq 1$ and also
such type of terms, e.g., $
\frac{(V_{n} -(n+m)\beta_{2})}{\Omega_{n}}\simeq 0$.
Therefore, the rescaled fluctuation factor for both, i.e. EHA and MEHA,
are almost similar except the ${Re}\langle \hat{a}^{2}(T)\rangle$ in
the MEHA involves $\cos (2T)$ additionally.
However, for particular values of the
 initial mean-photon number
 the maxima of $\cos(2T)$
occur in the course of  the revival times of $Q_{1}(T)$ and then the overall behavior does not affect.
The final remark, in the strong-intensity regime and
 when $\beta_{1}=\beta_{2}=\lambda=\omega$  the
 fluctuation factors of the MEHA--defined in the framework of  the slowly varying
operators--would be
typically as those of the EHA defined  in (\ref{14}).

From the discussion given in this section we can conclude that
generally the EHA can be used to investigate RCP for natural phenomenon approach but
 it is inadequate for numerical simulation approach.
Nevertheless,
for particular types of initial states--those for which
EHA and MEHA provide  almost similar behavior--EHA is adequate also for numerical simulation approach.

\section{Revival-collapse phenomenon in the amplitude-squared
fluctuations}
As we did in the previous  section we discuss briefly here whether the
higher-order fluctuation factors can carry  information on the corresponding
 atomic inversion or not.
As an example we consider the amplitude-squared fluctuations \cite{hil}.
The amplitude-squared fluctuations can occur in the fundamental mode in the second
harmonic generation and can be converted into normal fluctuations.
The two quadratures  correspond to the real and imaginary parts of the
 square of the field  amplitude  are
\begin{equation}
\hat{X}_{2}=\frac{1}{4}[\hat{a}^{2}+\hat{a}^{\dagger 2}], \quad
\hat{Y}_{2}=\frac{1}{4i}[\hat{a}^{2}-\hat{a}^{\dagger 2}]. \label{30}
\end{equation}
These quadratures obey the uncertainty relation
\begin{equation}
\left[\hat{X}_{2},\hat{Y}_{2}\right]=\frac{1}{4i}(2\hat{a}^{\dagger } \hat{a}+1). \label{30a}
\end{equation}
After minor calculation one can show that
the two  fluctuation factors associated with the amplitude-squared fluctuations are
\begin{eqnarray}
\begin{array}{lr}
F_{2}(T)=
 \langle\hat{a}^{\dagger 2}(T)\hat{a}^{2}(T)\rangle+{\rm Re}
\langle\hat{a}^{4}(T)\rangle-2({\rm
Re}\langle\hat{a}^{2}(T)\rangle)^{2},
\\
\\
S_{2}(T)=
 \langle\hat{a}^{\dagger 2}(T)\hat{a}^{2}(T)\rangle-{\rm Re}
\langle\hat{a}^{4}(T)\rangle-2({\rm
Im}\langle\hat{a}^{2}(T)\rangle)^{2},
\label{31}
\end{array}
\end{eqnarray}
it is said that  the system is able to yield amplitude-squared
fluctuation when $F_{2}(T)<0$ or $S_{2}(T)<0$.
Similar to section 3  we consider two approaches, which are natural phenomenon
and numerical simulation. These will be discussed in the following.
As the comparison between EHA and MEHA leads to
 conclusions similar to
those given in section 3 we will not discuss this issue in the
present section.
\subsection{Natural phenomenon}
In this part  we are seeking states, which
evolve with standard JCM, say, in such a way that the contribution of the
moments $\langle \hat{a}^{2}(T)\rangle$ and
$\langle \hat{a}^{4}(T)\rangle$  to the fluctuation factors
(\ref{31}) are negligible in the course of
the interaction. For such  states expressions (\ref{31}) reduce to
\begin{equation}
F_{2}(T)=S_{2}(T)=
 \langle\hat{a}^{\dagger 2}(T)\hat{a}^{2}(T)\rangle.
\label{31a}
\end{equation}
In fact, the quantity
 $\langle\hat{a}^{\dagger 2}(T)\hat{a}^{2}(T)\rangle$
can provide behavior  similar to that associated with the mean-photon
number, i.e. atomic inversion. We have already introduced a class of states,
which can fulfill the above requirements.  That is the $k$-photon coherent
states given by (\ref{1}) for  $k=3,5,7,..$ and
 the probability amplitudes are real.
Here we give some details about  the evolution of the
$3$rd-photon coherent states with the standard JCM when the atom is
 initially in the excited atomic state.
For this case one can easily show that
\begin{equation}
F_{2}(T)= \langle\hat{n}(0)\rangle^{2}-
\langle\hat{n}(0)\rangle \sum\limits_{n=0}^{\infty}
P(n)\cos(2T\sqrt{3n+4}), \label{33}
\end{equation}
where $\langle\hat{n}(0)\rangle$ is the initial mean-photon number of the
$3$rd-photon coherent state.
On the other hand, the corresponding atomic inversion is
\begin{equation}
\langle\hat{\sigma}_{z}(T)\rangle=
 \sum\limits_{n=0}^{\infty}
P(n)\cos(2T\sqrt{3n+1}). \label{34}
\end{equation}
In the strong-intensity regime expressions (\ref{33}) and (\ref{34}) yield

\begin{equation}
\langle\hat{\sigma}_{z}(T)\rangle\simeq
\frac{\langle\hat{n}(0)\rangle^{2}-F_{2}(T)}{\langle\hat{n}(0)\rangle}.
\label{35}
\end{equation}
Argument similar to that given for (\ref{21}) shows that the revival
time of the present case
can be obtained through  the relation
\begin{equation}
2 T_{R}\left[ \sqrt{\langle \hat{n}(0)\rangle+3}
-\sqrt{\langle \hat{n}(0)\rangle}\right]\simeq 2\pi, \label{20ab}
\end{equation}
which leads to $T_{R}=\frac{2\pi}{3}\sqrt{\langle
\hat{n}(0)\rangle}$, i.e. it is three times smaller than that
associated with   the initial coherent state case.
\subsection{Numerical simulation}
 Similar arguments
as those given in section 3 show that
amplitude-squared fluctuation factors (\ref{31}) of the $m$th ($m>2$) JCM can exhibit
behavior similar to that of $\langle\hat{\sigma}_{z}(T)\rangle_{m=1}$
only when the values  of  ${\rm Re}\langle\hat{a}^{2}(T)\rangle$ and ${\rm
 Im}\langle\hat{a}^{2}(T)\rangle$ are very small (or zeros). In this case
the forms of ${\rm Re}\langle\hat{a}^{4}(T)\rangle$  and
$\langle\hat{\sigma}_{z}(T)\rangle_{m=1}$ have to be comparable.
Using similar  procedures as those given in section 3
one can deduce the proportionality
factor as
\begin{equation}
f(n)=mn^{\frac{m-3}{2}}. \label{36}
\end{equation}
Expression (\ref{36}) indicates that $S_{2}(T)$
can provide  behavior
similar to that of $\langle\hat{\sigma}_{z}(T)\rangle_{m=1}$
 only when $m=3$.
 This  is similar to that associated with the normal fluctuation but
here $f(n)=3$. One can deduce the corresponding  rescaled amplitude-squared
fluctuation factor of the $3$rd JCM case, which includes
 behavior typical to that of $\langle\hat{\sigma}_{z}(T)\rangle_{m=1}$ is
\begin{equation}
Q_{2}(T)=\frac{S_{2}(\frac{1}{3}T)-\langle\hat{n}(0)\rangle^{2}}
{\langle\hat{n}(0)\rangle^{2}}.\label{37}
\end{equation}
Comparison between (\ref{29}) and (\ref{37}) shows that
the interaction time in (\ref{29})
is two times greater than that in (\ref{37}) owing to the fact that
  we  deal with the square of the field  amplitude.
Finally, similar to the normal-fluctuation case
the rescaled amplitude-squared fluctuation factor
(\ref{37}) is insensitive to the values
of the relative  phases of the atomic system.

\section{Conclusions}
In the present paper we have discussed the possibility of relating
the  information involved in the fluctuation factors of the $m$th JCM  to the atomic inversion of EHA.
We have made some comments on the differences between EHA and MEHA.
Generally, we have shown that there
are two approaches, namely, natural phenomenon and numerical simulation.
For the natural-phenomenon approach we have shown that there
is a class of states for which  fluctuation factors  can include information on the
corresponding atomic inversion naturally.
This has been shown
not only for normal fluctuations but also for amplitude-squared fluctuations.
Furthermore, for such approach fluctuation factors can exhibit coherent trapping based on
 the values of the relative phases of the atomic system.
On the other hand, for the numerical-simulation approach
 we have shown that for specific value of  $m$,
in particular $m=3$, the fluctuation factors (or one of them)
of the normal fluctuations as well as the amplitude-squared fluctuations can include
 RCP similar to that associated with  the
atomic inversion of the standard JCM. More illustratively,
the evolution of the quadrature fluctuations of an initially given
field state interacting with a two-level system by a three-photon
transition reflects the RCP phenomenon of the hypothetical interaction
of the same field state with a two-level system by one-photon
interaction where the level spacing is one third of that of the former system.
Furthermore, we have deduced the forms of the rescaled fluctuation
factors for this case,  which can involve
typical information on the atomic inversion of the standard JCM.
These forms would be helpful for  experimentalists.
In contrast to the natural approach fluctuation factors here are
insensitive to the values of the relative phases of the atomic system.

In fact these results are novel and indicate that
the homodyne detector \cite{homo} can be used to measure RCP.
In this respect the signal
coming from the microwave cavity is optically mixed with a strong coherent local oscillator
using 50:50 beam splitter. Then the emerging fields are detected and
the photocurrents are electronically treated in such a way that the measured
quantity is the rescaled fluctuation factors.
Quite recently similar setup is given for
measurement induced and quantum computation with atoms in optical
cavities \cite{anders}.
Moreover, in cavity QED, the homodyne detector
technique has been applied for the single Rydberg atom and one-photon field
aiming to study  the evolution of
the field phase for the regular  JCM \cite{haroc}.
Nevertheless, for the nonlinear version of the JCM in an ideal cavity ($Q=\infty$),
 e.g. two-photon JCM, the detuning parameter $\triangle$ should be
 much greater than the Rabi frequencies of the one-photon transition
($\triangle=33.3 MHz$ in Cs, $\triangle=39 MHz$ in $^{85}$Rb);
thus the the Stark shift and the two-photon coupling are appreciable
\cite{puri}. Moreover, the progress in the trapped ions \cite{Lei} and micromaser
\cite{mas1}
are promising to produce the phenomenon presented in this paper.
This is related to the fact that the two-photon Rydberg atom has been already
realized in the micromaser \cite{mas2}. We hope in the near future that it would be possible to
produce a frequency within the range allowed by the equation $\omega=
\omega_1+\omega_2+\omega_3$ where $\hbar \omega$ is the energy difference
between the two levels and $\omega_1, \omega_2, \omega_3$ are the frequencies
of the three photons generated by the transition.

\section*{\bf Acknowledgments}
I would like to thank deeply Professor Sarge Haroche (Coll\'ege de France, 11 place Marcelin Berthelot, F-75231 Paris Cedex 05, France)
for drawing my attention to the references [27].
I thank Professor A.-S. Obada
 (Department of Mathematics, Faculty of Natural Science,
Al-Azhar University,
Nasr City, Cairo, Egypt) for the critical reading of the original paper.
Also I am grateful to the International
Islamic University Malaysia for hospitality and financial support.
\section*{References}

\end{document}